\documentstyle[epsf]{l-aa}
\newcommand{\be}{\begin{equation}}
\newcommand{\ee}{\end{equation}}
\newcommand{\bd}{\begin{displaymath}}
\newcommand{\ed}{\end{displaymath}}
\newcommand{\gapprox}{\;\rlap{\lower 2.5pt                              
             \hbox{$\sim$}}\raise 1.5pt\hbox{$>$}\;}                    
\newcommand{\lapprox}{\;\rlap{\lower 2.5pt                              
             \hbox{$\sim$}}\raise 1.5pt\hbox{$<$}\;}

\newcommand{\rd}{{\rm d}}   
\begin{document}
   \thesaurus{06    % A&A Section 6: Form. struct. and evolut. of stars
              02.07.2 ; %GW
              02.01.2 ; %accretion
              02.09.1 ; %instabilities
              02.13.1 ; %B-fields
              08.14.1 ; %N*
              13.07.1 ; %GRB
              }
   \title{Gamma-ray bursts from X-ray binaries}
%   \subtitle{}

% 
   \author{H.C. Spruit \inst{1}}

%   \offprints{}

   \institute{ Max-Planck-Institut f\"ur Astrophysik, Postfach 1523, 
               D-85740 Garching bei M\"unchen, Germany}

   \date{Received 2 November, accepted 4 November 1998}

   \maketitle

%   \maintitlerunninghead{}

%   \authorrunninghead{}

\begin{abstract}

A weakly magnetized ($\sim 10^7$G) neutron star, slowly spun up by 
accretion in an X-ray binary, crosses the instability boundary for r-mode 
instability at P=1--2 msec. The amplitude of the oscillation, which initially 
increases only at the spinup time scale, is secularly unstable due to the negative 
temperature dependence of the viscosity in neutron star matter, and diverges 
after a few hundred years. Angular momentum loss  by the gravitational wave 
causes strong differential rotation, in which the magnetic field is wound up to 
$10^{17}$G on a time scale of a few months. When this field becomes unstable to 
buoyancy instability, a surface field strength of a few $10^{16}$G is produced on 
a time scale of seconds, which then powers a GRB with energies of 
$10^{51}$--$10^{52}$ and duration of 1--100 sec.

\keywords{gamma-ray bursts -- X-ray binaries -- gravitational waves --
magnetic fields -- instabilities}

\end{abstract}

\section{Introduction}
A significant difficulty with several proposed mechanisms for gamma-ray
bursts (GRB, Fishman and Meegan 1995) is that only a small amount of baryons 
($\lapprox 10^{-5}$M$_\odot$) can be involved if the observed relativistic expansion 
speeds are to be produced 
(Shemi and Piran 1990, Paczy\'nski 1990). Scenarios involving the coalescence of 
two neutron stars (Narayan, Paczy\'nski and Piran 1992, Ruffert and Janka, 
1998), or special types of supernovae (Paczy\'nski 1998, MacFadyen and Woosley 
1998) have been proposed, but both involve potentially large amounts of 
`contaminating' baryons. A baryon-poor mechanism is the rapid spindown of a 
pulsar with a rotation period on the order of a millisecond and a magnetic field 
strength of $10^{16-17}$G (Usov 1992). The problem with this idea is finding a 
convincing scenario for magnetizing a millisecond pulsar sufficiently rapidly to 
such a field strength (Klu\'zniak and Ruderman 1998, Dai and Lu 1998). 

A neutron star with a mass of 1.4 M$_\odot$, a radius of 10 km and a rotation 
period of a millisecond contains a rotational energy of $2\times 10^{52}$ erg, the right 
amount to power a gamma-ray burst. Known neutron stars with such periods are 
the milliscond pulsars, believed to be descendants of X-ray binaries such as Sco 
X-1 or SAX J1808.4-3658 (P=2.5 msec, Wijnands and van der Klis 1998). 
Accretion of angular momentum from the mass transfering secondary has slowly 
spun up the neutron star in such a system over a period of $10^8$--$10^9$yr. 
When the secondary has been evaporated (Ruderman et al. 1989, Tavani and London 
1993), a single millisecond pulsar remains. Its rotation energy is slowly 
dissipated by pulsar radiation. If the magnetic dipole moment of the star is 
$\mu$, the electromagnetic torque braking these `cosmic flywheels' is (Shapiro 
and Teukolsky 1983).
\be T_{\rm EM}={2\over 3} {\mu^2\Omega^3 \over c^3},\label{TEM} \ee
where $\Omega$ is the star's rotation rate. If $B_{\rm s}=10^8B_8$G  is the 
star's surface field strength, and $P=10^{-3}P_{-3}$ s 
its rotation period, the spin-down time scale by this torque is $4\times 
10^9\,{\rm yr}{P_{-3}^2/ B_8^2}$. Extraction of the rotation energy on the short 
time scale of a GRB  requires a much higher field strength, of the order 
$10^{16}$G. If the star rotates differentially by an amount 
$\Delta\Omega/\Omega\sim 1$, field strengths up to $10^{17}$G are produced by 
winding-up of an initially weak field. Such fields are also strong enough to 
overcome the stable density stratification in the star, and float to the stellar 
surface (Klu\'zniak and Ruderman 1998) to produce dipole fields of the order 
$10^{16}$G. The problem of making a GRB from an X-ray binary is thus reduced to 
finding a plausible way to make the star rotate differentially. 

\section{Runaway gravitational wave instability}
Rotating neutron stars are naturally unstable to excitation of their oscillation 
modes by the emission of gravitational waves (Shapiro and Teukolsky 1983). A 
mode is excited like a squeaking brake. The deformation of the oscillation star 
provides surface roughness and the gravitational wave the torque coupling the 
star to the vacuum outside, analogous to the frictional coupling of a brake lining. 
Rotational modes (Rossby waves) turn out to be much more effectively excited 
than the pulsation modes of the star (Andersson 1997, Andersson et al. 1998). 
Modes with angular quantum numbers $l=m=2$ grow fastest. The unstable 
displacements in these modes are mainly along horizontal surfaces. Their 
frequency in an inertial frame is $2/3\, \Omega$, and in the corotating frame 
$-4/3\, \Omega$ (retrograde). In the absence of viscous damping, the growth rate of the most unstable mode is (Andersson et al. 1998)
\be \sigma_{\rm GW}=5\times 10^{-2} P_{-3}^{-6}\quad{\rm s}^{-1}.\ee
Including viscous damping, the growth rate is
\be \sigma=\sigma_{\rm GW}-1/\tau_{\rm v}, \ee
where $\tau_{\rm v}$ is the viscous damping time $\tau_{\rm v}=(\Delta 
R)^2/\nu$, and $\Delta R\approx 0.5R$ is the width of the unstable eigenfunction. 
For the kinematic viscosity $\nu$ the standard value quoted (van Riper 1991) for 
a neutron superfluid is $\nu\approx 2\times 10^7T_7^{-2}$ cm$^2$s$^{-1}$ at an 
average density of $3\times 10^{14}$ g cm$^{-3}$, where $T=10^7T_7$ is the temperature.  This expression holds at low temperatures; at temperatures above $\sim 3\times 10^9$K the viscosity starts increasing strongly with $T$. During the spinup by 
accretion, the temperature of the star is determined by the accretion rate, 
$T_7\approx {\dot M}_{-9}^{1/4}$, where $\dot M=10^{-9}\dot M_{-9}$ 
M$_\odot$/yr is the mass accretion rate. With the typical value $\dot M_{-9}=1$, 
the critical rotation period for instability to set in is 3 msec. 

SAX J1808.4-3658 has a period shorter than this, and millisecond radio pulsars are 
known with periods as short as 1.5msec. This suggests that the instability actually sets 
in at somewhat shorter periods. This is possible if the standard value 
underestimates the viscosity somewhat. Assuming that this is the case, I will, in 
the following, allow the viscosity to be increased, by a constant factor $f_{\rm 
v}$. For example, with  $f_{\rm v}=30$ the critical period is 1.5msec at an 
accretion rate $\dot M_{-9}=1$.

\begin{figure}
\mbox{}\hfill\epsfysize5cm\epsfbox{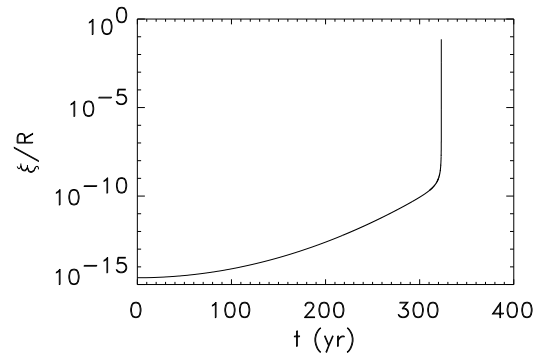}\hfill\mbox{}
\caption{\label{runaway}
Growth of the r-mode oscillation amplitude of a neutron star due to gravitational
wave emission. Time in years since onset of instability, amplitude is the
fractional displacement amplitude. The final runaway is due to the low viscosity
of the interior, as it is heated by viscous dissipation of the mode}
\end{figure}

A star slowly spun up by accretion (on a time scale of $10^9$yr) would become 
weakly unstable to the gravitational wave mechanism as its period decreases 
below the critical value. The unstable mode dissipates energy by viscous friction. 
Together with the negative temperature dependence of the viscosity this makes 
the mode amplitude secularly unstable, leading to a runaway of the mode 
amplitude in a finite time (this has been noted independently by Levin, 1998). The mode energy $E$ is governed by 
\be \rd E/\rd t=\sigma E,\label{emode}\ee
The rotation rate of the star, assumed uniform at this stage, is determined by the 
accretion torque $T_{\rm a}$ and the gravitational wave angular momentum loss 
$T_{\rm GW}$:
\be I\rd\Omega/\rd t=T_{\rm a}-T_{\rm GW}=(GMR)^{1/2}\dot M-{3\over 
2}{\sigma_{\rm GW}\over\Omega}E. \ee
The heating is determined by viscous dissipation of the mode:
\be {\rd U\over \rd t}={\nu\over(\Delta R)^2}E-L_\nu, \ee
where $U\approx 3\times 10^{43}T_7^2$ is the internal energy integrated over 
the star (van Riper 1991). $L_\nu\approx 10^{40}T_{9}^8$ represents the loss by 
neutrino cooling (van Riper 1991), and becomes important only at high 
temperatures. The energy losses by radiation at the surface can be neglected on 
the time scale of interest. The solution of these equations is shown in Figure 
\ref{runaway}. A few hundred years after the rotation rate has crossed the 
instability threshold, the mode amplitude diverges. During the final runaway 
phase the temperature is so high that viscous damping is weak and the mode 
grows on a time scale of the order of minutes. 

\section{Differential rotation and winding-up of the magnetic field}
The angular momentum loss by gravitational wave emission varies through the 
star, depending on the shape of the unstable eigenfunction. This loss does not 
match the angular momentum distribution of a uniformly rotating star, so 
differential rotation is induced. This winds the initial magnetic field into a 
toroidal (azimuthal) field. Because the rotation period is so short, the 
nonaxisymmetric components of the initial field are quickly eliminated by 
magnetic diffusion (R\"adler 1981) in a process similar to `convective expulsion' 
of magnetic fields (Weiss 1966); the remaining axisymmetric and nearly 
azimuthal field is as sketched in Klu\'zniak and Ruderman (1998). To represent 
these processes in a simple model, I divide the star into two zones with boundary 
at $r\approx 0.5R$, such that the outer shell has approximately the same moment 
of inertia as the core, and rotating at rates $\Omega_{\rm s}$ and $\Omega_{\rm 
c}$ respectively. The angular momentum loss by gravitational waves is taken to 
be restricted to the core for simplicity. If the initial radial field component is 
$B_r$, the azimuthal field component increases as
\be \rd B_\phi/\rd t=(\Omega_{\rm s}-\Omega_{\rm c}) B_{r}. \ee
Between core and shell the magnetic torque $T_{\rm m}={2\over 3}r^3B_rB_\phi$ 
acts, opposing the differential rotation. 

\begin{figure}
\vbox{
\hbox{\mbox{}\hfill\epsfysize5cm\epsfbox{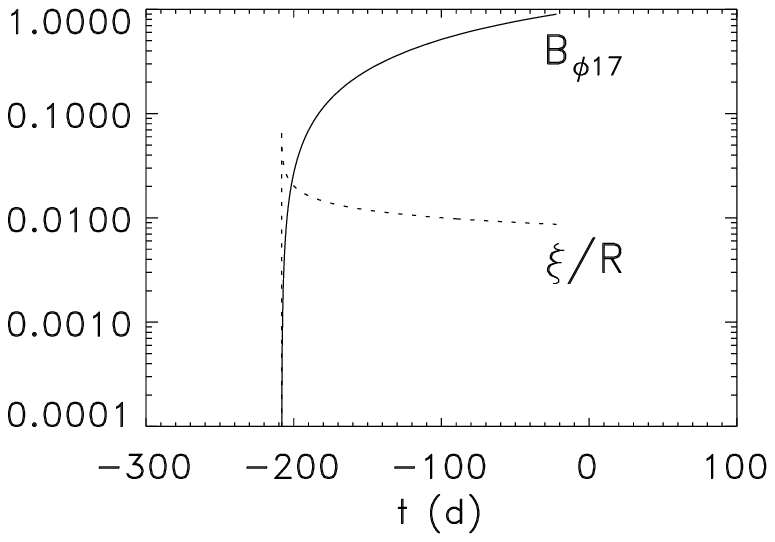}\hfill\mbox{}}
\hbox{\mbox{}\hfill\epsfysize5cm\epsfbox{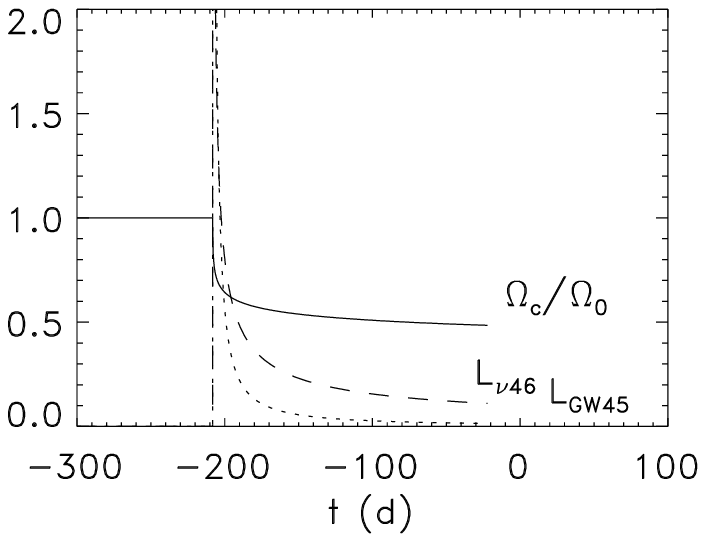}\hfill\mbox{}}}
\caption{\label{wind} {\bf top}: Winding up of the magnetic field of the star 
(solid, in units of $10^{17}$G) by differential rotation in the saturated phase of 
the gravitational wave instability. This phase of the process ends (at $t=0$) 
when the magnetic field strength exceeds the critical value for buoyant 
instability of the field. {\bf bottom}: Core rotation rate $\Omega_{\rm c}$ 
(relative to its initial value $\Omega_0$, solid), gravitational wave (dotted) and 
neutrino (dashed)
luminosities (in units of $10^{44}$ and $10^{45}$ erg/s, respectively)}
\end{figure}

The growth of the unstable mode is limited at large amplitude by nonlinear mode 
couplings. Since these have not been computed yet, I parametrize them by adding 
a term $-\sigma_{\rm GW}E^2/E_{\rm s}$ to the right hand side of (\ref{emode}), where $E_{\rm 
s}$ is the level at which the mode saturates. The displacement amplitude of the 
mode at this level is $\xi_{\rm s}\approx (E_{\rm s}/M)^{1/2}/\Omega$. The 
evolution of the azimuthal field strength, core rotation rate, gravitational wave 
luminosity and neutrino luminosity during the runaway phase is shown in figure 
\ref{wind} for a case with an assumed saturation level $\xi/R=0.03$. The 
luminosities peak when the mode first reaches this level. After this, the mode 
amplitude decreases as the core spins down by the gravitational wave torque. 

\section{Buoyant instability of the toroidal field}
The increasing azimuthal magnetic field inside the star does not manifest itself 
at the surface until it becomes unstable to buoyancy instability. For this to 
happen the field  has to become strong enough (Klu\'zniak and Ruderman 
1998), of the order $B_{\rm c}=10^{17}$G, to overcome the stable stratification 
associated with the increasing neutron/proton ratio with depth. The growth time 
$\tau_{\rm B}$ of the instability is of the order of the Alfv\'en travel time, 
$\tau_{\rm B}\sim R/(V_{\rm A}-V_{\rm Ac})$, where $V_{\rm Ac}$ is the 
Alfv\'en speed at the critical field strength $B_{\rm c}$. Buoyancy instability is 
generally nonaxisymmetric, forming loops of field lines which erupt through the 
surface (e.g.\ Matsumoto and Shibata, 1992). The action of the radial velocities in 
the eruption process on the azimuthal field produces a strong radial field 
component, which couples to the differential rotation. The whole process is 
likely to yield a complex time dependent field configuration, beyond the scope of 
this investigation, but important aspects can be illustrated with a simple model. 
Let $B_r$ be a representative value for the amplitude of the radial field 
component developing in the star. Its growth is due to the buoyant instability, but 
limited to a value of the order of the azimuthal field strength:
\begin{eqnarray} \rd B_r/\rd t=&(v_{\rm A}-v_{\rm Ac})(B_\phi^2-B_r^2)^{1/2}/R, \qquad (v_{\rm A}>v_{\rm Ac}) \cr
=&0 \qquad (v_{\rm A}<v_{\rm Ac}). \label{ebr}
\end{eqnarray}
The radial field component appearing at the surface will have a complicated 
structure, of which only the lower multipole components contribute to the field 
at the light cylinder, where the electromagnetic energy emission starts. To take 
this into account, I assume an effective dipole field strength at the surface, 
$B_{\rm s}=\epsilon B_r$, with $\epsilon<1$. Including the electromagnetic 
torque (\ref{TEM}) due to this surface field, the rotation rates of the core and 
shell are given by
\be I_{\rm s}\rd \Omega_{\rm s}/\rd t= -T_{\rm m}-T_{\rm EM}, \ee
\be I_{\rm c}\rd \Omega_{\rm c}/\rd t= T_{\rm m}-T_{\rm GW}. \label{rc}\ee

\begin{figure}[t]
\vbox{
\hbox{\mbox{}\hfill\epsfysize5cm\epsfbox{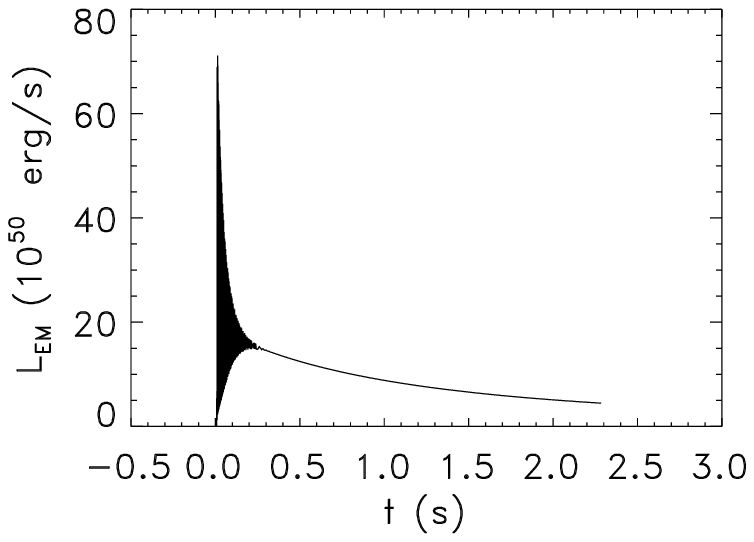}\hfill\mbox{}}
\hbox{\mbox{}\hfill\epsfysize5cm\epsfbox{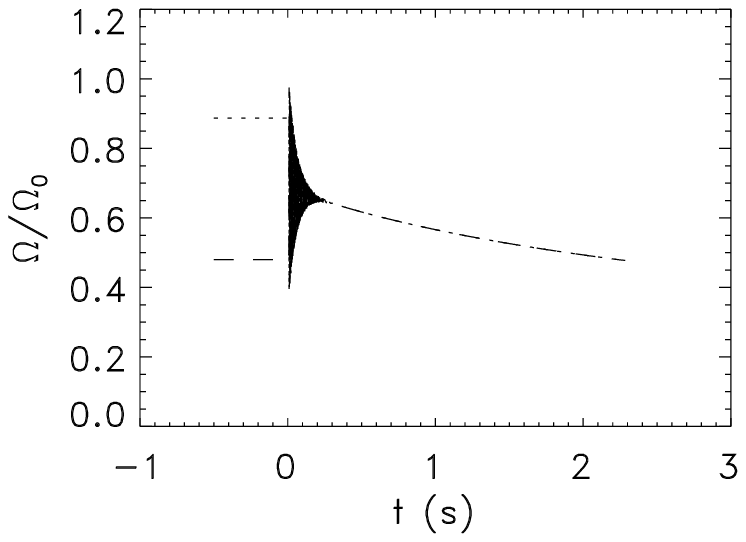}\hfill\mbox{}}}
\caption{\label{burst} {\bf top}:  gamma-ray burst luminosity (in units of 
$10^{50}$ erg/s). The ringing is an Alfv\'enic oscillation due to the coupling 
between core and shell by the strong radial field component. An ad hoc damping 
term has been added to represent magnetic dissipation inside the star, making 
the ringing decay during the burst. {\bf bottom}: evolution of the rotation rates of 
shell (dotted) and core (dashed) during the burst. The strong radial field quickly 
couples the two}
\end{figure}

The solution of eqs.~(\ref{ebr}--\ref{rc}) starting at the time when the critical 
field strength $B_{\rm c}$ is reached, is shown in fig.~\ref{burst} for a case 
with $\epsilon=0.3$. The final rise of the burst is on an Alfv\'en time scale 
$\tau_{\rm A}=R/v_{\rm A}\approx 0.5$ msec. The ringing seen in the light curve 
is an Alfv\'enic oscillation, in which the shell and the core exchange angular 
momentum by the magnetic torque $T_{\rm m}$. In this simple model, it is a 
regular oscillation; in reality, the angular momentum exchange between different 
parts of the star will lead to a much more complicated variation, but on 
comparable time scales. Observations show clear evidence for such sub-msec 
variability in the central engine of GRB (e.g. Bhat et al. 1992, Walker et al. 1998).  

Some 70\% of the initial rotation energy of the star is released during the burst. 
The remaining 30\% is released in the form of gravitational waves and neutrinos, 
mostly during the peak of the gravitational wave instability (fig 2) a few months 
before the electromagnetic burst. 

\section{Discussion}
A number of parameters are involved in the caculations presented, in particular 
the initial field strength $B_0$ of the star, the assumed saturation amplitude 
$\xi_{\rm s}$ of the r-mode oscillation, the critical field strength for buoyant 
instability $B_{\rm c}$ and the viscosity factor $f_{\rm v}$. Variation of $B_0$ 
has little effect on the duration and amplitude of the burst; it mainly affects the duration of the winding-up phase (fig. \ref{wind}), which decreases with 
increasing $B_0$. The saturation amplitude also has little effect, and also 
influences mainly the duration of the winding-up phase (decreasing with 
$\xi_{\rm s}$). The value of $B_{\rm c}$ is more critical: if it is taken higher 
than about $3\times  10^{17}$G, there is not enough energy in the differential 
rotation to reach buoyant instability. On the other hand, lowering $B_{\rm c}$ has 
little effect on the burst duration and amplitude; as buoyant instability sets in, 
the field is quickly amplified to a few times $10^{17}$G by the differential 
rotation, and the rest of the burst is the same as before. 

The most important parameter is the initial rotation rate $\Omega_0$, i.e. the 
rate at the time when r-mode instability sets in. This rate is determined by the 
viscosity at that time, which depends on the assumed viscosity factor $f_{\rm 
v}$ and, through the temperature dependence of the viscosity, on the accretion 
rate (actually, the average of $\dot M$ over a thermal time scale preceding the 
instability). The total energy of the burst increases as $\Omega_0^2$, the 
duration of the burst decreases as $\Omega_0^{-2}$. The peak burst luminosity, 
which is critical for the detection of GRB, therefore increases as $\Omega_0^4$. 
In the above, I have tuned $f_{\rm v}$ to get a critical rotation period of 
1.5msec, a value suggested by the known periods of msec pulsars. In the present 
scheme, these would be the remnants of X-ray binaries that managed to escape 
becoming GRB. Theoretical work to determine if such a high viscosity factor is 
realistic would be desirable, however. 

At the distance of the secondary star ($\sim 10^{11}$cm), the radiation energy 
density during the burst is of the same order as the binding energy density of the 
secondary. It is therefore likely that a significant fraction of the 
secondary's  mass is stripped in the process, or that it is destroyed altogether. 
% The slow 
%expansion (compared with the relativistic expansion of the burst itself) of 
%this mass may explain the optical afterglow observed in some GRB, as in 
%the model of Blinnikov and Postnov (1998). 

Since the life time of an X-ray binary ($\sim 10^8-10^9$yr) is modest on a 
cosmological scale,  the GRB rate predicted by the model would roughly 
scale as the cosmological star formation rate. X-ray binaries are observed in the 
galactic disk, so that the model puts GRB in galactic disks as well, as observed. 
They would not be closely related to individual star forming regions (cf. 
Paczy\'nski, 1998), however, since the life time of these regions is less than the 
age of X-ray binaries. The neutron star binary merging scenario is more likely to 
produce GRBs somewhat outside galaxies, since the typical kick of 300 km/s received 
by the last-formed neutron star would have caused such binaries to travel $\sim$ 
100 kpc before merging. (X-ray binaries are a special population that must have 
received small kicks,  since typical ones would have unbound these binaries). 

After the burst there remains a slowly rotating hot neutron star with a (dipole) 
field strength of a few $10^{16}$G. Due to its violent origin, this field 
configuration is likely to be unstable and rearrange itself on various time scales, 
including ones much longer than that of the burst. This predicts that most bursts 
would show low-level activity continuing after the main burst, probably decaying as 
a power law of time. The rate of decay would depend on the decay rate of the field 
strength. If this decay last sufficiently long, the remnant would behave like a soft 
gamma repeater, as in the model of Thompson and Duncan (1995). Perhaps some of the 
current soft gamma repeaters and anomalous X-ray  pulsars (van Paradijs et al. 1995, 
Mereghetti et al. 1998, Kouveliotou et al. 1998), with present field strengths of 
$\sim 3\times 10^{14}$G, may actually be the remnants of GRB that took place in our 
galaxy $10^5$--$10^6$ years ago.

\begin{acknowledgements}
I thank the anonymous referee for his competent review, for suggesting additions,  
and for his very speedy reply. I also thank Andrew Ulmer for useful comments and 
suggestions for improvement.
\end{acknowledgements}

\end{document}